\documentclass[11pt]{article}
\usepackage[pctex32]{graphics}
\usepackage{amssymb}
\usepackage{latexsym}
\usepackage{epsfig} 
\usepackage{multirow} 
\usepackage{pstcol}
\usepackage{graphicx}
\usepackage{float}
\usepackage{epstopdf}
\usepackage{mathrsfs}
\usepackage{subcaption}

\usepackage{amsmath}
\usepackage{amssymb}
\usepackage{relsize}
\usepackage{multirow}

\usepackage{graphicx}
\usepackage{multicol} 

\usepackage{times}
\usepackage{epstopdf}

\definecolor{Red}{rgb}{1,0.,0.}
\usepackage{fancyhdr}
\usepackage{setspace}
\title{Metapopulation network models for understanding, predicting and managing the coronavirus disease COVID-19}
\author{D Calvetti$^{1,2}$ \and A Hoover$^3$ \and J Rose$^2$ \and E Somersalo$^{1,2}$}
\date{$^1$ Department of Mathematics, Applied Mathematics, and Statistics\\
Case Western Reserve University\\
$^2$ Center for Community Health Integration\\
Case Western Reserve University\\
$^3$ Department of Mathematics \\
University of Akron}

\begin{document}
\maketitle

\begin{abstract}

%%% Leave the Abstract empty if your article does not require one, please see the Summary Table for full details.
Mathematical models of SARS-CoV-2 (the virus which causes COVID-19) spread are used for guiding the design of mitigation steps
aimed at containing and decelerating the contagion, and at identifying impending breaches of health
care system surge capacity. The challenges of having only lacunary
information about daily new infections and mortality counts are compounded by the geographic
heterogeneity of the population. This combination complicates prediction,
particularly when homogenized population models with an underlying assumption of well mixed
cohorts are used. To  address this problem, we propose to account for the differences between
rural and urban settings using network-based, distributed models where the spread of the pandemic
is described in distinct local cohorts with nested SEIR models. The setting of the model parameters
takes into account the fact that SARS-CoV-2 transmission occurs mostly via human-to-human
contact, and that the frequency of contact among individuals differs between 
urban and rural areas, and may change over time. Moreover, the probability that the virus spreads
into an uninfected community is associated with influx of individuals from other communities
where the infection is present. To account for these important aspects, each node of the network
is characterized by the frequency of contact between its members and by its level of connectivity
with other nodes. Census and cell phone data can be used to set up the adjacency matrix of the
network, which can, in turn, be modified to account for different levels of mitigation measures.
In order to make the network SEIR model that we propose easy to customize, it is formulated
in terms of easily interpretable parameters that can be estimated from available community level data. 
The models parameters are estimated with Bayesian techniques using COVID-19 data for the states of Ohio and Michigan. 
The network model also gives rise to a geographically distributed
computational model that explains the geographic dynamics of the contagion, e.g., in larger cities
surrounded by suburban and rural areas.

\end{abstract}

\section{Introduction}
Predicting the spread of COVID-19 is critical to public health decision making, including decisions to relax mitigation measures in different communities. Data on the number of individuals testing positive for the novel Corona virus in every county of the USA is updated continuously and used to inform mathematical models for predicting how the pandemic will evolve. The novelty of the virus and the current lack of testing capacity for the general population add to the challenges of the task, and can explain the wide variability seen in model predictions.

It has been established that the virus is transmitted through respiratory droplets and that symptoms range from non-existent to life-threatening.  The range of clinical presentation has also been wide and includes constitutional, respiratory, gastrointestinal, dermatologic, and musculoskeletal signs and symptoms. Unlike the case for SARS (Severe Acute Respiratory Syndrome), where the virus transmission appears to have occurred primarily after the emergence of symptoms,
there is evidence that viral shedding occurs, and may even peak, during the few days just prior to symptom onset \cite{He2020}.	
The social distancing recommendation to keep at least 6 ft away form other individuals and to wear face masks in public places are addressing the concerns about asymptomatic
infections, as are state-level bans on large gatherings.

The dynamics of COVID-19 spread suggest that once the virus enters densely populated communities with no mitigation measures in place, the disease is likely to flare up rapidly, infecting a large number of individuals in a short amount of time, with the danger of overwhelming the local health system, as has been the case in the region around Milan, Italy and in New York City. Not surprisingly, the first COVID-19 outbreaks in many countries have been recorded in cities that, being major economic hubs or tourist centers, have had significant contact with previously infected areas. Contact between communities enables COVID-19 spread, and needs to be taken into consideration when predicting where the next hot spots are likely to occur.

The dynamics of the spread of COVID-19 shows that once the virus enters densely populated communities with no mitigation measures in place, the disease is likely to flare up rapidly, infecting a large number of individuals in a short amount of time, with the danger of overwhelming the local health system, as has been the case in the region around Milan, Italy and in New York City. Not surprisingly, the first COVID-19 infection in many countries has been recorded in cities that, being major economic hubs or tourist centers, have had lot of contacts with previously infected areas. The rate of contacts between communities is responsible for the spread of COVID-19, and needs to be taken into consideration when predicting where the next hots spots are likely to be next.

Classical mathematical models for the spread of epidemics subdivide the population into three cohorts of susceptible (S), infected (I) and Recovered (R) individuals, (classical SIR models of \cite{Kermack}), with the possible addition of a fourth Exposed (E) cohort, accounting for the incubation time before infection begins, (SEIR models). Both SIR and SEIR models assume that the underlying population and the subpopulations within each compartment are well mixed, a necessary condition for the mean field description to be accurate. In the case of COVID-19, the lack of immunity of the population due to the novelty of the virus and the ease of transmission make its spread very sensitive to the type and frequency of contacts among individuals in the community. Such contact depends partly on population density; therefore separately modeling communities with differing population density can reduce violation of the mean field assumption.

In addition to different contact rates, traffic between communities also plays an important role in the spread of the pandemic, as individuals arriving from an area with a large number of infections act as potential vectors for the virus previously uninfected communities. Although homogenized SIR and SEIR models are not suited to account for these important aspects of the spread of COVID-19, they can form the fundamental units of a metapopulation model of interconnected communities, Separate sets of parameters account for community-specific settings. In addition to providing a more realistic explanation of the geographic pattern of spread, metapopulation models can also be used to test which changes in commuting patterns are more likely to keep the pandemic under control.

Finally, it is important that COVID-19 models acknowledge the time dependency of model parameters. The number and type of contacts change in response both to mitigation measures imposed and to a population?s awareness of risks and ongoing adherence to public health guidance. The time course of key parameters can provide valuable insight into the spread patterns and aggressiveness of the disease, as well as into the effectiveness of various mitigation measures.

In this time when different states in the USA are debating whether relaxing mitigation strategies and travel restrictions are likely to create new hot spots in areas little affected by the pandemic, a predictive model that can be adapted to different regions can immediate applicability. In response to these needs, the aim of this study is to adapt a new network SEIR inspired model of COVID-19 spread to understand the dynamics of the pandemic in a network of 18 counties in the region of Northeast Ohio around Cleveland, and a network of 19 counties in Southeast Michigan around Detroit. The computational model addresses  all of the points discussed above: The model parameters may be variable in time, and up to date Bayesian computational tools are used to inform the model on a daily basis regarding the progression of the epidemic, providing also an estimate of the uncertainties of the estimates. The metapopulation model uses census data to track population density and the movement of individuals between communities \cite{ACS2015}. Furthermore, the model gives an estimate of the size of the asymptomatic cohort based on the observed new infection count. Finally, to avoid introducing intractable parameters that limit use and interpretation, the basic model is kept as simple as possible without overlooking some fundamental characteristics of COVID-19. The output comprises interpretable quantities that can be immediately communicated to public health and health system decision makers, and allow a comparison with existing model predictions. The computational framework can be shown to reproduce within reasonable uncertainty the observed spreading timeline between the communities included in this study; and the predictive skills over relatively short time windows, up to a few weeks, can be demonstrated.

 \section{Materials and Methods}
 
 \subsection{Predictive models of COVID-19}
 
In the absence of data from previous outbreaks, mathematical and statistical models \cite{Murray2020} are the main tool to predict how the COVID-19 outbreak will develop, including estimates of how many patients will need to be hospitalized, the expected number of admissions to ICU and the type of resources that the health systems should have ready. As the situation is dynamic, computational methods capable of dynamic model updating as data is gathered are of key importance. 
 
 \subsection{A SE(A)IR model of COVID-19 spread}
 
In epidemiology, a standard mathematical tool to describe how an infectious disease spreads through a population is an SIR model, describing how the changes in time of the three cohorts of susceptible ($S$), infected ($I$) and recovered ($R$) individuals. The model can be modified to account for an incubation period during which an individual has been exposed but has not yet become infectious by adding the exposed ($E$) cohort. This is the essence of the SEIR model of \cite{Kermack} which has been used successfully to model epidemics for nearly a century. In the case of the COVID-19, there is evidence of exposed individuals shedding the virus already a few days before developing any symptoms. In fact, according to some recent laboratory tests the amount of virus released is largest right before the onset of the symptoms, and over the next few days it start to decay. Moreover, antibodies found in individuals who did not report any symptoms of COVID-19, points towards a presence of a potentially large number of asymptomatic infectious individuals, who spread the virus without being detected, due to the current practice of reserving the testing for persons with clear symptoms. Furthermore, if the $I$ cohort in the SEIR model comprises symptomatic infected and infectious individuals who have tested positive for COVID-19, it is reasonable to assume that most of them will be in some form of isolation, 
hence with limited contribution to the spreading of the infection.

\begin{figure}[h!]
\begin{center}
{\includegraphics{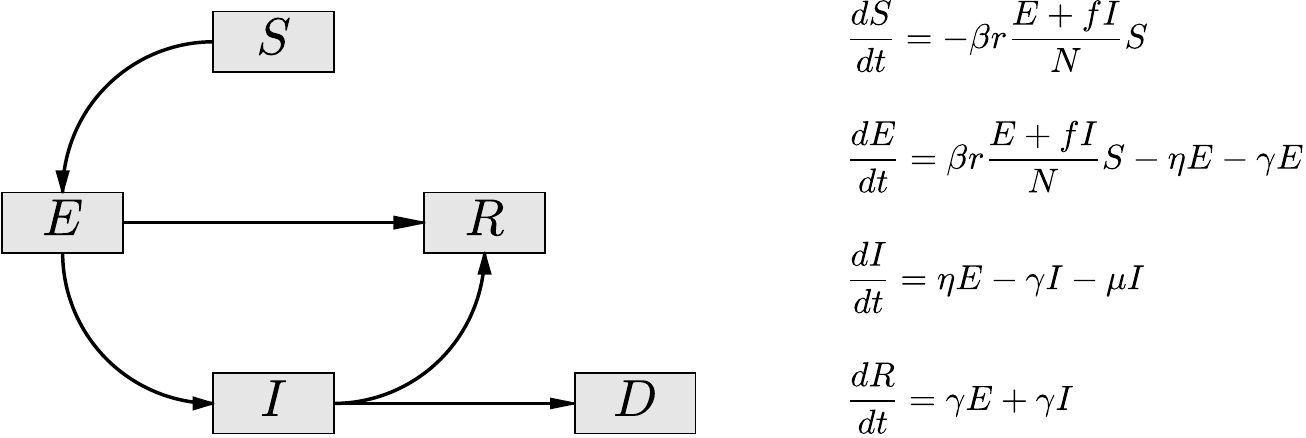}}
\end{center}
\caption{\label{fig:SE(A)IR} The compartment diagram of the SE(A)IR model. Compared to the standard SEIR model, the flux $E\rightarrow R$ has been added, and the nonlinear interaction term is modified by the replacement $I\rightarrow E+fI$. Here, $0<f<1$ account for the fact that diagnosed symptomatic individuals are in partial isolation, contributing less to the infection than the asymptomatic individuals ($A$) immersed in the exposed cohort $E$ of the model. In the presented calculations, the value $f=0.1$ is used.}
\end{figure}

To account for the possibility that the asymptomatic/oligosymptomatic infectious pool is mostly responsible for the spread of the pandemic, we modify the basic SEIR model by interpreting the cohort $E$ as ``exposed, infective, and asymptomatic''  a portion of which will recover without becoming symptomatic, and the cohort mostly responsible for infecting the susceptible population, with a smaller contribution from the infected cohort. To underline this interpretative difference, and the immersion of the asymptomatic ($A$) cohort in $E$, we refer to the model as SE(A)IR model. Figure~\ref{fig:SE(A)IR} shows a schematic compartment diagram and the governing equations, illustrating the modification to the standard SEIR model to account for asymptomatic infectious individuals. 

% FIGURE 1

When used for interpretation of real data, the classical compartment models usually suffer from two limitations: The model parameters are constant in time, while the population interaction is a dynamic process, and (ii) the models operate under the hypothesis that the population is well-mixed, while in reality the data are an aggregate of underlying sub-populations with a complex interaction structure. Both of these limitations have been acknowledged and addressed in the literature (see \cite{XX}). Below is the description of our contribution and findings, with an emphasis on two particular cases, population models of Northeastern Ohio and Southeastern Michigan in the period  from early March 2020 to early May 2020.

 \subsection{Bayesian estimation of the SE(A)IR parameters}
 
The governing equations of the SE(A)IR model (see Figure~\ref{fig:SE(A)IR}) depend on a number of parameters reflecting the characteristics of the epidemic: more precisely, $\beta>0$ is the infectivity rate, or probability of contagion, $r$ the number of contacts per day of infectious and susceptible cohorts, $\eta = 1/T_{\rm inc}$ the incubation rate in days, the reciprocal to the expected time of incubation of the disease, $\gamma = 1/T_{\rm rec}$ the expected recovery rate, with $T_{\rm rec}$ the  expected number of recovery days, and $\mu$ the mortality rate. 
The latter three parameters (incubation, recovery, and mortality rates) are strongly pathogen dependent and to some extent sensitive to factors like demographics and co-morbidities, but for the modeling purposes, they can be considered as independent of time. Arguably, the most important parameter is the product $\beta r$, controlling the rate at which susceptible individuals are infected. While the value of $\beta$ depends mostly on the infectivity power of the virus, the factor $r$ accounting for the frequency of contacts between infectious and susceptible individuals and may change significantly with the introduction or lifting of social distancing. In fact, the effectiveness of the measures can be directly monitored by estimating this quantity, and in particular, a time dependency of it. 

In the methodology that we propose, for each subpopulation of individual counties, the product $r \beta$ is assumed to be time dependent, and is estimated using a Bayesian filtering technique known as particle filtering (PF), discussed by \cite{liu2001combined,arnold2014parameter,arnold2013linear,arnold2015astrocytic}. IN PF, thousands of realizations  (particles) of the model, each with its own set of parameter values and cohort sizes, simulate the day-by-day propagation of the epidemics. Each day, the predictions of the particles for the next day are computed and compared to the new data, and particles whose results are in better agreement with the data are retained and replicated, while particles less well fit to explain the data are discarded. After this ``survival of the fittest'' step, the replicated particles are proliferated through a randomization, guaranteeing a rich variability of the particle cloud to account for the variations in the next time step. Since the effects of changing mitigation strategies are reflected in the value of $r \beta$, this quantity is updated daily, thus providing a time series for each particle. As pointed out, unlike in standard epidemiology model, we do not assume that the quantity is constant.

The data used of estimating the parameters and the current cohort sizes comprises the daily count of confirmed new infections $I$, while no direct data of the cohort size of asymptomatic and exposed $E$ is assumed to be available. Therefore, estimating the time evolution of the state vector $(S,E,I,R)$ together with the parameter $r\beta$ provides direct information about the number of asymptomatic individuals. One of the key parameters of interest to us is the ratio $\rho = E/I$. It turns out that this ratio tends towards an equilibrium value, $\rho\rightarrow \rho^*$, as the infection progresses, allowing us to define in a very natural way an equivalent of the {\em basic reproduction number} $R_0$ of the SE(A)IR model.  In the classical SIR model, the basic reproduction number is defined as a dimensionless quantity $R_0 = (r\beta)/(\gamma+\mu)$, and a wealth of literature exists for generalization and estimation  of $R_0$ for more complex models, see, \cite{Delamater,Diekmann,diekmann2010construction,dietz1993estimation,heesterbeek2002brief,Heffernan}, as well as some critical views on its usefulness, as in \cite{li2011failure}. For the current model, the equivalent form is
\[
 R_0 = \rho^*\frac{\eta}{\gamma +\mu},
\]
and it turns out that the equilibrium value  $\rho^*$ can be estimated in a straightforward manner if an estimate for the product $\beta r$ is available, see Figure~\ref{fig:R0 vs beta r} for further clarification of the symbols. The novel $R_0$ has a similar role in the model as it has in the SIR model, that is, the infections spreads only if $R_0>1$, thus giving a useful summary for policy makers of the success of the mitigation efforts. Conversely, the above formula provides a means of estimating $\rho^*$ and thereby the size of the asymptomatic cohort if $R_0$ has been estimated from the data, e.g., by fitting an exponential to the cumulative data of infected individuals.

One of the advantages of the particle filtering approach over data fitting approaches is that it allows us to assess the model uncertainties. At each time step, thousands of realizations of every quantity of interest is computed, and of those realizations, one can generate histograms, posterior intervals of different degrees of belief, expectations, and median values. In particular, the time traces of the quantities are not summarized in a single curve but are presented as posterior envelopes, or credible envelopes of given level of belief. Moreover, the particles can be propagated in the future to provide predictive envelopes of future data.

\subsection{Metapopulation network models of COVID-19 spread}

The travel of individuals carrying the virus between communities is the main engine for spreading pandemics, both a the local and a global level. The pattern followed by the spread of COVID-19, similar to that of the 1918 influenza, indicates that typically at first the flair-up occurs in larger cities  with a high population density, moving to smaller, more rural communities when the number of new infected in the cities has already decreased. The predictions of mathematical models for the COVID-19 spread in a network of diverse connected communities can be used to understand where the next hot spots are likely to move, and to design mitigation measures to keep the epidemic from overwhelming the healthcare system. 

As pointed out in the literature \cite{wang2014spatial}, well mixed compartment models have a limited capability of explaining the dynamics of the epidemics in large heterogenous populations, as they ignore the local dynamics depending on population density, segregation of diverse groups, and geographic separation  of communities.  In line with the county by county reporting of the cases, {\em metapopulation network models} (MNM) are an appropriate tool to address the population inhomogeneity. To model the effect of daily commuter traffic on COVID-19 spread, an MNM can be designed as a directed graph where the counties constitute the nodes and the weights of the directed edges are proportional to the number of commuters between pairs of counties. Although the epidemic within each county can be described locally with a SE(A)IR model, the model cohorts need to be adjusted to reflect the movement of individuals between counties, thus affecting the infection dynamics. 

% FIGURE 2
\begin{figure}[h!]
\begin{center}
{\includegraphics{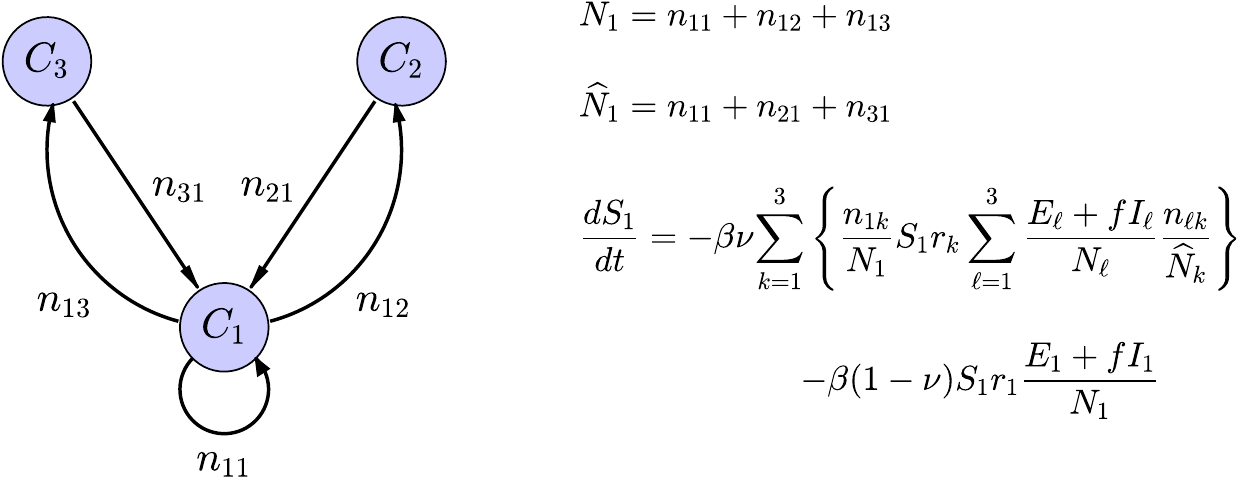}}
\end{center}
\caption{\label{fig:network SEIR} Schematic representation of the network model with three nodes, with only the in-links and out-links of $C_1$ included. The symbol $n_{jk}$ indicates the 
number of residents of node $C_j$ commuting to node $C_k$, the home community being an option, and $N_k$ is the population size resident in node $C_k$, while $\widehat N_k$ is the size of the population working at the community $C_k$. The average number of daily contacts in community $C_k$ is $r_k$, and $\eta$ is the fraction of time spent in the destination of commute, and $1-\eta$ is the fraction of time spent in the home community. The formula for the rate of change of $S_1$ only is shown. The two contagion terms correspond to infections through contacts that the susceptible population $S_1$ resident in $C_1$ obtain in the commuting destinations and the ones obtained in home community.}
\end{figure}
 
The interaction between the subpopulations can be built in the interaction term of the SE(A)IR model, see Figure~\ref{fig:network SEIR} for an explanation. Infections in a given node  arise through contacts between susceptible residents of the node with infectious individuals in target nodes of the commuting traffic, the domiciliary node included, plus the infections that happen in the home community, e.g., during weekends and evenings. These two infection mechanisms are included in the model with weights proportional to the average time spent in work/leisure outside the home community and the time spent at home.  Each community has its own characteristic number $r_j$ of daily contacts, the number being presumably higher in densely populated urban centers than in sparsely populated rural communities with fewer interaction opportunities. The value of $r_j$ will decrease in response to the adoption and adherence to mitigation measures that discourage group gatherings.

\subsection{Two network models: Northeastern Ohio and Southeastern Michigan}

The methodology was tested with the daily updated infection data corresponding to 18 counties in Northeastern Ohio, listed in Table S1 in the Supplementary Material, and with 19 counties in Southeastern Michigan, see Table S2 in the Supplementary Material. The comparison of the two regions is of particular interest, as they both represent a population with similar cultural background and to some extent similar demographics and mixtures of dense urban areas, suburban commuter communities and rural areas. However, the mitigation measures were introduced slightly differently, and at a different level of the epidemics. 	
  
Commuter data was procured from the Census Bureau's American Community Survey (ACS) of 2015. In the ACS, commuter data for the residents of a county was compiled that quantified the number of residents leaving this county for work in another county. The focus of this study was on the 18 county region that comprises Northeast Ohio and the 19 county region the comprises Southeastern Ohio, therefore the commuting data was used for commuter traffic within these regions. This is integrated into our model by computing $n_{jk}$  from the data from the ACS of the number of people commuting from county $j$ to county $k$. Commuting data for counties outside of the region of interest is not used in this model, but the region of interests were chosen the envelop the urban centers of Northeast Ohio, (e.g. Cleveland, Akron, Canton, and Youngstown) and Southeast Michigan (e.g. Detroit, Flint, Ann Arbor, and Lansing).
   
\section{Results}

Before reporting the results with the metapopulation network model, we present some of the preliminary results obtained with particle filter for each individual county. This information captures the characteristics of the transmission within a community prior to accounting for the contribution from the mobility. A discussion of the computational details can be found in the manuscript \cite{ParameterEstimation}. 
\subsection{Parameter estimation in individual communities}

\begin{figure}[h!]
\begin{center}
{\includegraphics{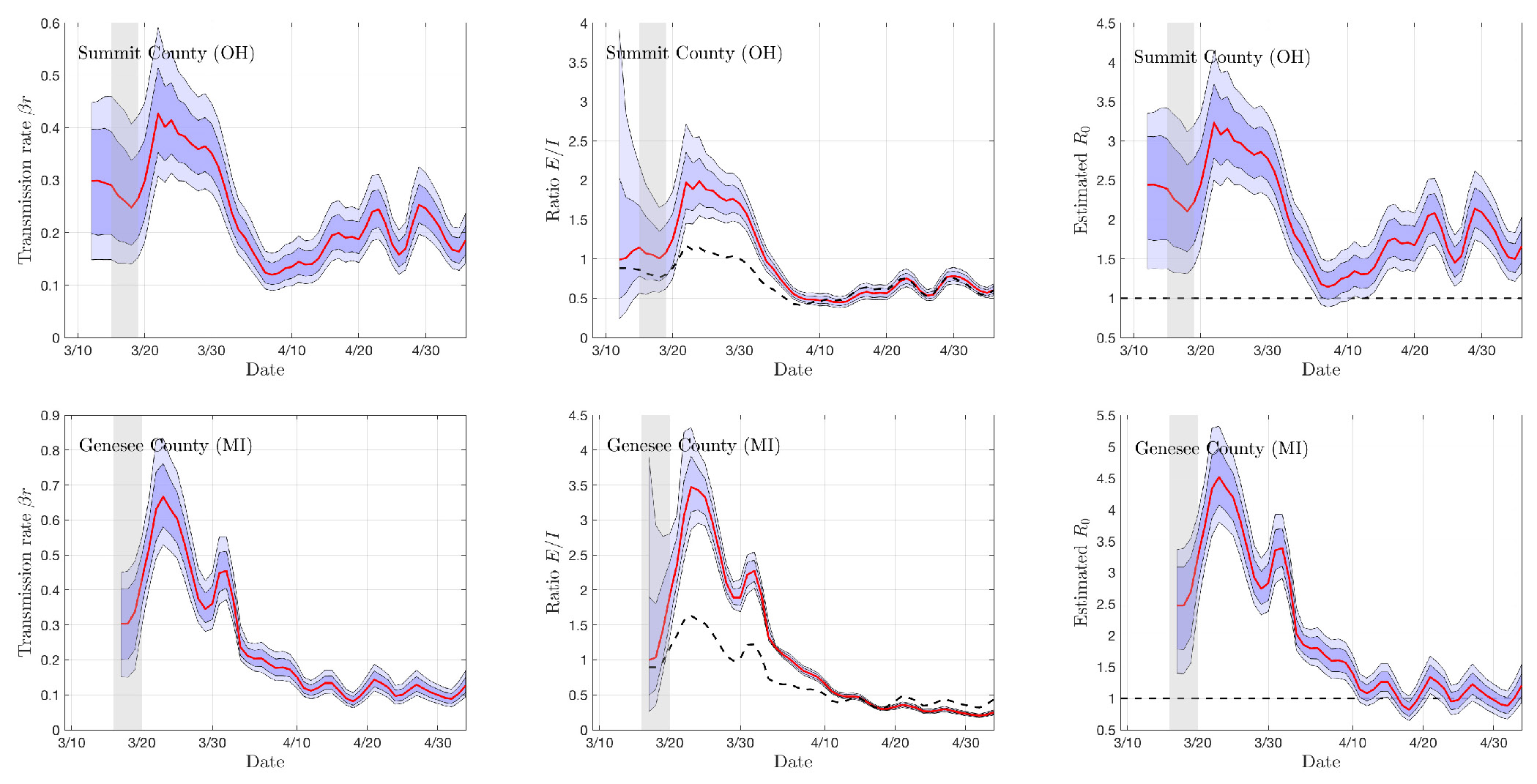}}
\end{center}
\caption{\label{fig:param} Sample outputs of the PF algorithm. The top row shows results based on daily new confirmed positive cases in Summit County (OH), and the second row in Genesee County (MI). In each plot, the envelopes represent the 50\% (dark) and 75\% (light) posterior belief. 
From left to right: The estimate of the rate of contagion $\beta r$, the ratio of number of exposed (asymptomatic) individuals and the infected symptomatic individuals, and the estimated basic reproduction number $R_0$. In the middle column, the dashed curve represents the equilibrium value of the ratio, and in the right column, the horizontal dashed line is indicates the critical value $R_0=1$. The gray vertical shading indicates the dates when the respective state started the social distancing measures. Observe that the drop in $R_0$ and $r\beta$ appear with a lag of about a week.}
\end{figure}

\begin{figure}[h!]
\begin{center}
{\includegraphics{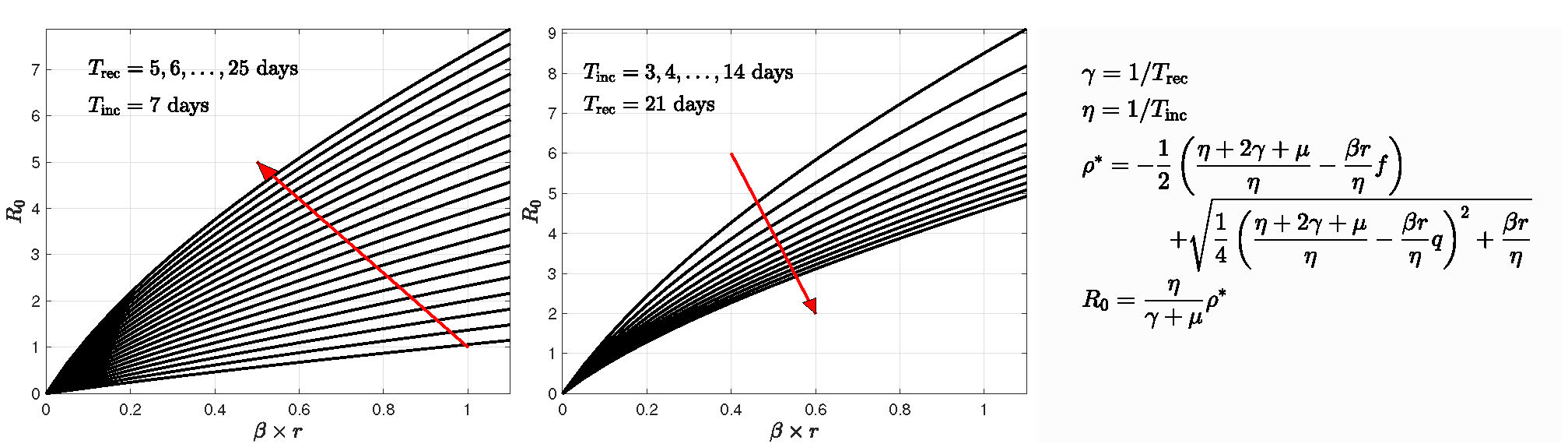}}
\end{center}
\caption{\label{fig:R0 vs beta r} The variation of $R_0$ as a function of the infectivity $\beta r$ with varying recovery time (left) and varying incubation time (right). The red arrows indicate the direction of growth of the time constants $T_{\rm rec}$ or $T_{\rm inc}$, respectively. The derivation of the formula for $R_0$ is given in \cite{ParameterEstimation}. In the form given here, it is implicitly assumed that the infection is in its outbreak phase, and no herd immunity is present.}
\end{figure}

To get a preliminary estimate of the transmission rate, in each individual county, we used the particle filter with the daily new infections data from all of the $18+19$ counties, and generated credibility envelopes for: The infection rate parameter $\beta r$, the ratio of the asymptomatic and infected cohort sizes, $\rho = E/I$, the $R_0$ derived from the estimated parameter $\beta r$ (see Figure~\ref{fig:R0 vs beta r}). Figure~\ref{fig:param} shows the outcomes for select counties. 

% FIGURE 3
\begin{figure}[h!]
\begin{center}
{\includegraphics{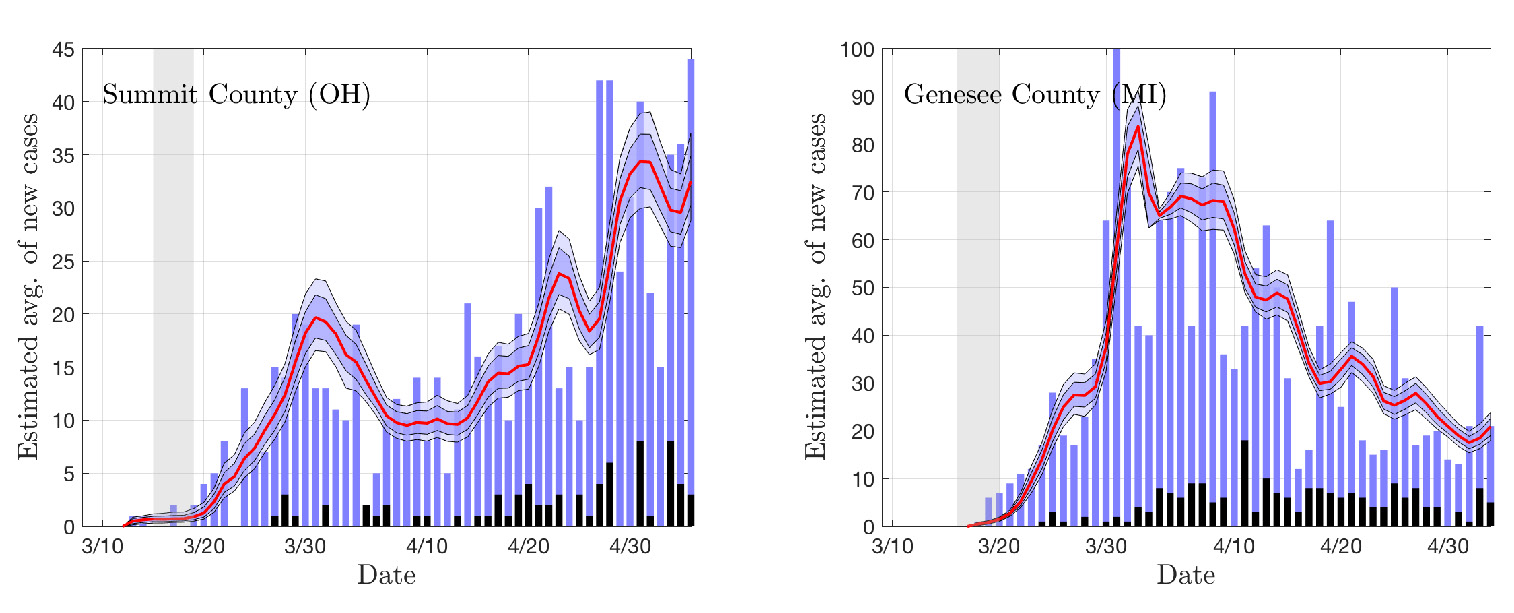}}
\end{center}
\caption{\label{fig:data} The daily new infection data from two counties, summit County in Ohio (Akron) and Genesee County in Michigan (Flint). The blue envelopes represent the 50\% (dark) and 75\% (light) posterior model uncertainty of the expected value of new infections, and the red curve represents the median. The dark gray columns are the number of deceased (not used in the estimation). 
The vertical gray shading indicates the period in which the state-wide mitigation measures were implemented (Ohio: 3/15/20-3/19/20, Michigan: 3/16/20-3/23/20). The number of particles is 5\,000.
In the model, the new infection count is assumed to be Poisson distributed around the expected value.}
\end{figure}

% FIGURE 4

In the sample calculations, the recovery and incubation times were kept constants, $T_{\rm rec} = 21$ days, $T_{\rm inc} = 7$ days. 
Figure~\ref{fig:R0 vs beta r} sheds some light on how the $R_0$ changes if different values were used. Here, the $R_0$ value is plotted against the rate of contagion with different combinations of the two time constants.

% FIGURE 5

% FIGURE 6
\begin{figure}[h!]
\begin{center}
{\includegraphics{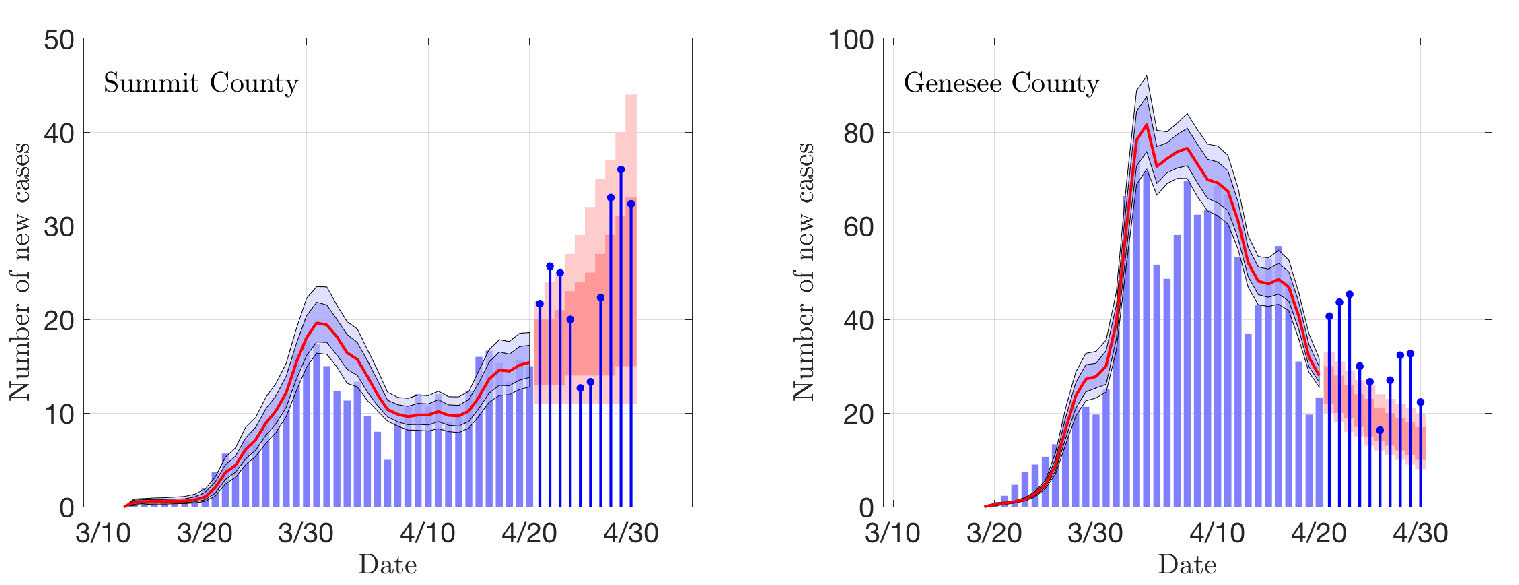}}
\end{center}
\caption{\label{fig:prediction} Examples of the prediction skill of the PF method. The PF updating is stopped ten days before the last data, and the particles are propagated using the last update of the states as initial values, and corresponding estimates for the parameter $\beta r$. The data are simulated drawing them from Poisson distribution, and the 50\% and 75\% predictive intervals for the data are computed. The true data is shown as stem plots. Observe that the predicted trend corresponds well to the estimated $R_0$.}
\end{figure}

Figure~\ref{fig:prediction} illustrates of the prediction skill of the PF method. In the figure, the last ten day's data are left out from the PF update and the state/parameter estimation is stopped early. Consequently, the state and parameter values for each particle are propagated forward for ten days, and the predicted average of the new cases for each particle is calculated. These average values are used as means for a Poisson process, and random realizations of predicted new cases for each particle is computed. Finally, the predicted data are used to calculate the predictive envelopes of given level of belief. In general, the true data may not fall in the predicted intervals, but in general, the algorithm anticipates the trend rather well. Observe that the predictions are not based on curve fitting, as the dynamics is determined by the full state vector containing components (susceptible and asymptomatic cohorts) that are not directly observed.  

 \subsection{Visualization of the predictions of the network model}

One of the central questions in the network modeling is how to trace the spreading of the infection between the nodes. To see if the network model is realistic, we estimate the delay of the onset of infection in the nodes after the infection is started in one of them. To validate the results with real data, the infection is initiated in the node with the first confirmed case: In Ohio, the first first infection was reported on 3/9/2020 in Cuyahoga County (Cleveland) and in Michigan on 3/10/2020 in Wayne County (Detroit).

\subsubsection{Reference case}
\label{sect:referencecase}

Using the estimated transmission rates for the two regions of interest, we are able to delineate differences in contact frequency $r$ between urban and rural counties. Contact frequencies for each county were chosen to be in line with the peak transmission rate, which presumably corresponds to the contact frequency before state issued stay-at-home orders were fully in place. The simulation was then initiated with one infected individual in Cuyahoga County and Wayne County, respectively, and run for 20 weeks.

\begin{figure}[h!]
\begin{center}
{\includegraphics[width=\textwidth]{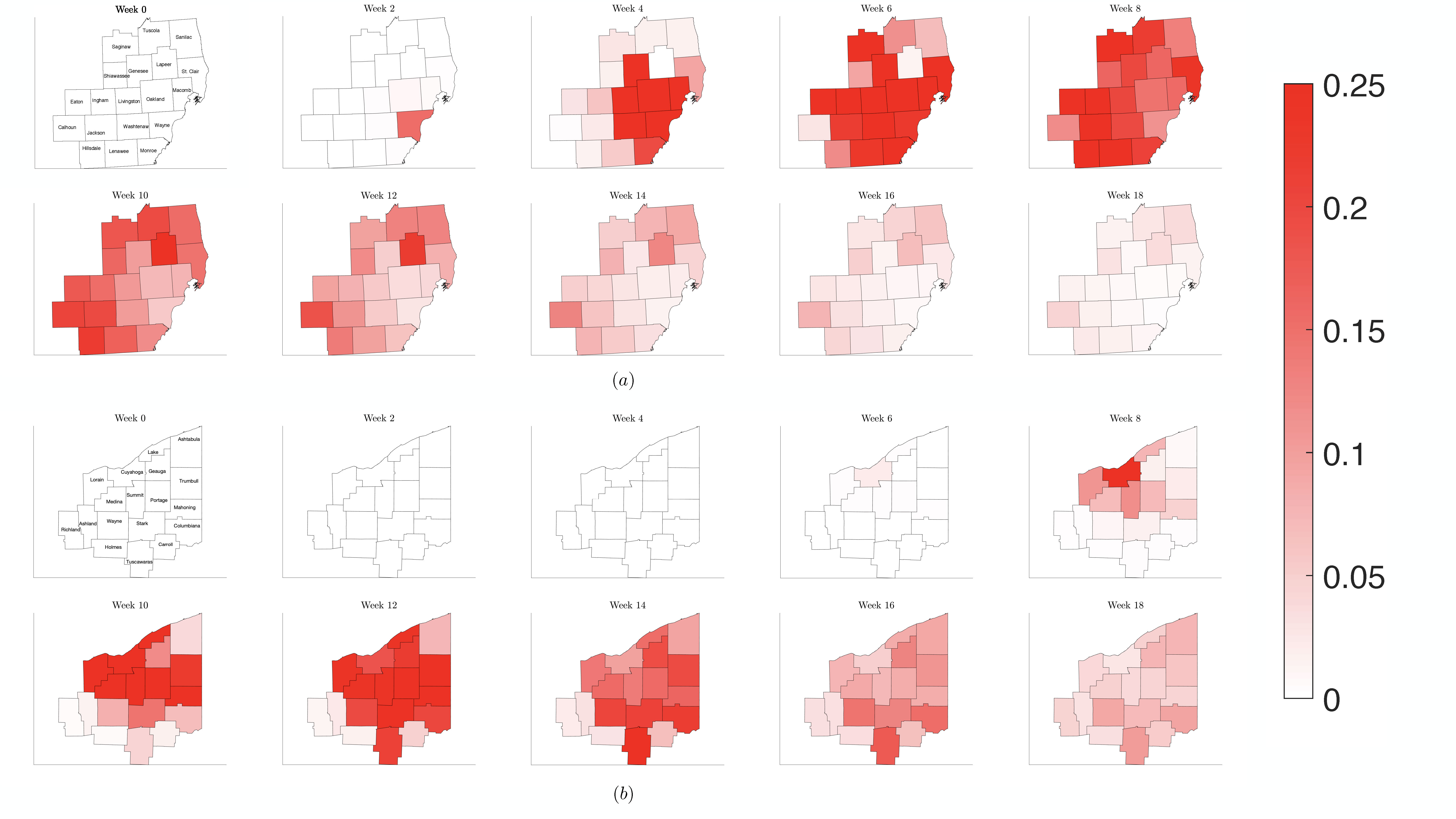}}
\end{center}
\caption{\label{fig:ref_basemap} Map of the counties in the region of interest of (a) southeast Michigan and (b) northeast Ohio, where the color on the county corresponds to the fraction of the population that is infected for the reference case of Sect.~\ref{sect:referencecase}.}
\end{figure}

Plotting the percentage of infected individuals relative to the population over a map of the counties of interest in Southeast Michigan and Northeast Ohio in Figure~\ref{fig:ref_basemap}, we are able to observe the dynamics of the spread of COVID-19 infections over the two regions. For Southeast Michigan, we note the initial rise in infections in Wayne County, as well as the two neighboring, densely populated communities of Oakland County and Macomb County, which form the greater Detroit metropolitan area. As the infection spreads in the Detroit metropolitan area, surrounding counties begin to see a rise in infection. However this spread does not correlate with physical proximity but rather commuter traffic, as seen with Lapeer County, a sparsely populated county that physically borders Macomb County and Oakland County but does not have an interstate connection to either, which experiences a spike in cases nearly 10 weeks after the initial infection in Wayne County. As the infection takes hold in each county, the number of infected peaks before recovering slowly, with differences in the relative peaks due to the differing populations of each county.

Observing the result of the northeast Ohio simulations in Figure~\ref{fig:ref_basemap}, we find that the spread of infection takes longer to fully realize in Cuyahoga County. This is in part due to the lower estimated transmission rates for the set of Ohio counties, reflecting a lower frequency of contacts. Once the infection has taken hold in Cuyahoga County, the infection then spreads to neighboring counties and grows in urban population centers in Mahoning, Stark, and Summit counties, as well as Lake and Lorain counties, which contain suburban bedroom communities that have highways that connect to Cuyahoga. Similarly to what was observed for southeast Michigan, we note that the number of infected peaks quickly before subsiding at slower rate.

\begin{figure}[h!]
\begin{center}
{\includegraphics[width=\textwidth]{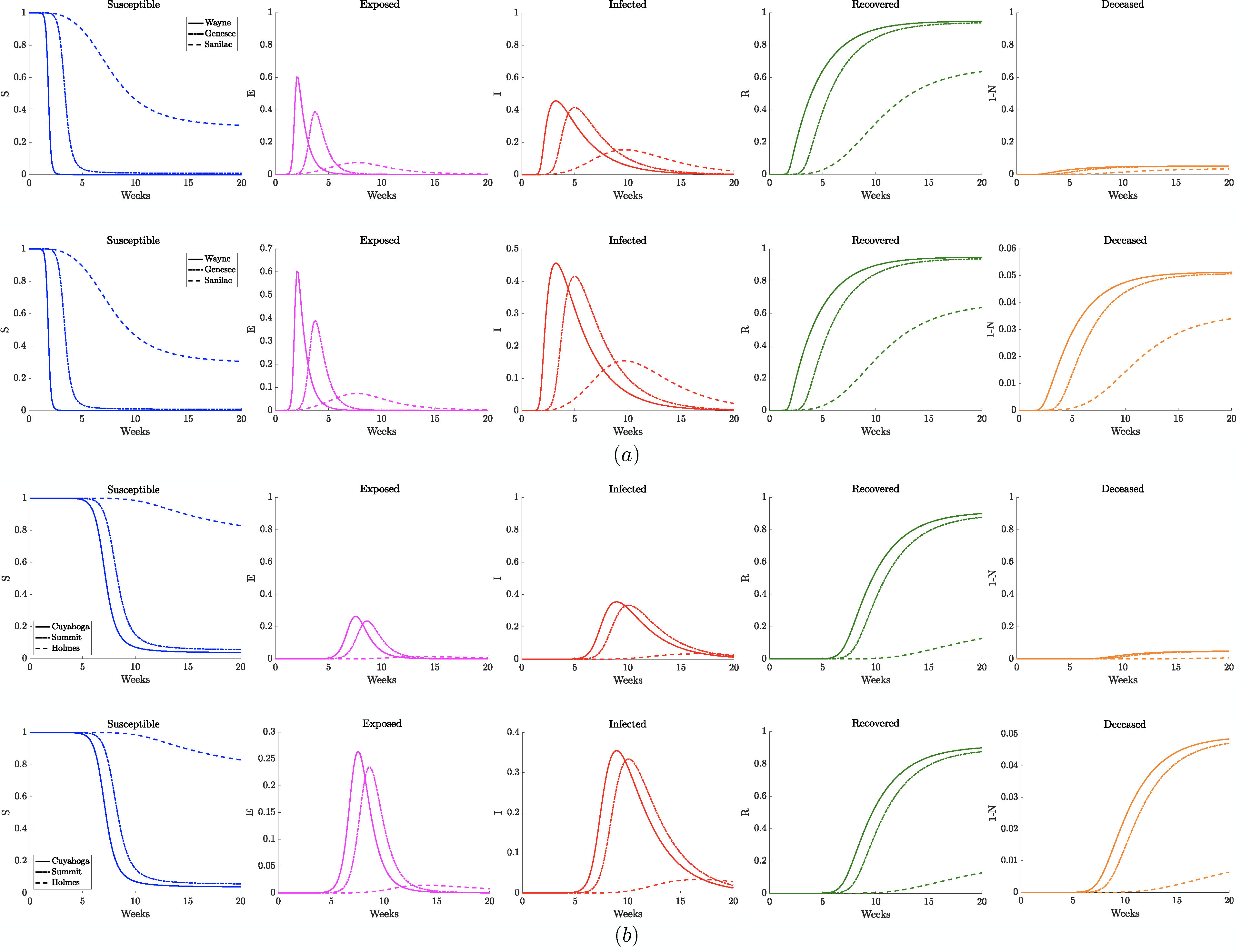}}
\end{center}
\caption{\label{fig:ref_SEIRD} Plots of the relative (left to right) susceptible, exposed, infected, recovered, and deceased populations for the reduced traffic study (Sect.~\ref{sect:referencecase}) for (a) Wayne, Genesee, and Sanilac counties in southeast Michigan and (b) Cuyahoga, Summit, and Holmes counties in northeast Ohio. The plots on top have the same y-axis scales and the plots on the bottom are with a relative scale for the y-axis.}
\end{figure}

In Figure~\ref{fig:ref_SEIRD} we plot the relative number of susceptible, exposed, infected, deceased, and recovered population for three counties of differing population densities for both Southeast Michigan and Northeast Ohio. In Michigan the focus is on Wayne County (high density, population of $1,257,584$), Genesee County (medium density, population of $405,813$), and Sanilac County (low density, population of $41,170$). We observe the initial sharp spike from Wayne County, followed by a spike in Genesee County. Sanilac County experiences its spike as the number of infected decreases in both Wayne and Genesee counties. Note that the relative peak in infections is higher in Wayne County and Genesee County than Sanilac County. This highlights the role of population density and contact frequency in the propagation of the epidemic.

In Ohio, we chose to examine Cuyahoga County (high density, population of $1,235,072$), Summit County (medium density, population of $541,013$), and Holmes County (low density, population of $43,960$), which correspond roughly to the same profile as Wayne, Genesee, and Sanilac counties, respectively. We observe that the relative peaks in infected population are staggered, with Cuyahoga County experiencing the first spike in infections, followed soon after by Summit County, and eventually by Holmes County. We also note that the height of the peak for relative number of infections is fairly similar for both Cuyahoga and Summit counties. That is in part due to the relatively similar contact frequency between the two counties, whereas Wayne County has a much higher contact frequency than Genesee County. Holmes County, which has lower contact rate and population, experiences a much more mild peak many weeks after the peaks in the other two counties.
  
% FIGURE 7 

% FIGURE 8 

\subsubsection{Reduced traffic}
\label{sect:redtraffic}

\begin{figure}[h!]
\begin{center}
{\includegraphics[width=\textwidth]{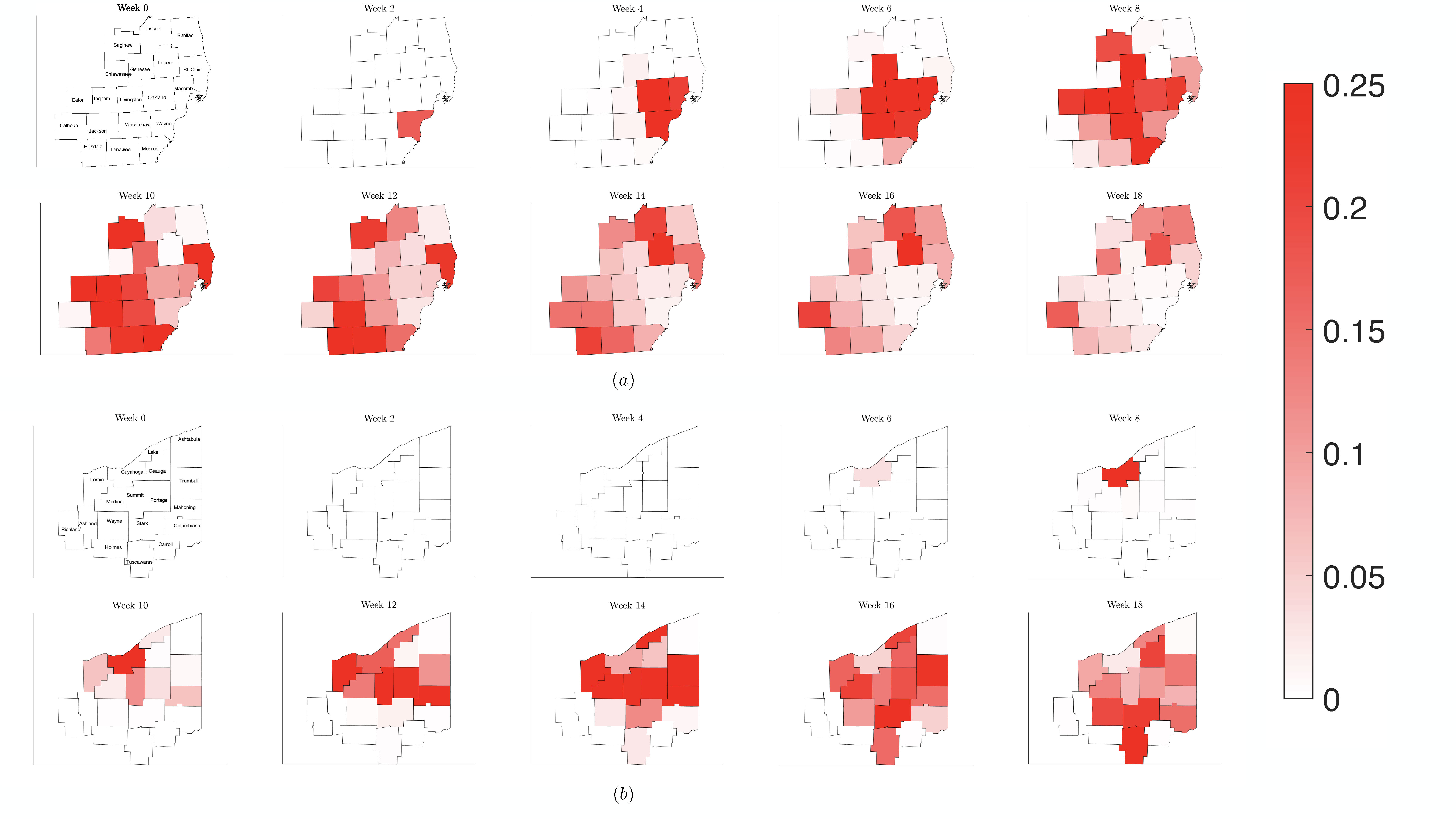}}
\end{center}
\caption{\label{fig:redtraffic_basemap} Map of the counties in the region of interest of (a) southeast Michigan and (b) northeast Ohio, where the color on the county corresponds to the fraction of the population that is infected for the reduced traffic study of Sect.~\ref{sect:redtraffic}.}
\end{figure}

In order the examine the role of commuting on the spread of the disease, we run a simulation reducing the number of commuters from other counties to 1\% of the original amount, while keeping the contact frequency unchanged inside each county. This would be akin to nearly shutting down each county but allowing people to continue to move about in their county. Plotting the relative population of infected on the counties of each region in Figure~\ref{fig:redtraffic_basemap}, we find  a similar picture to the reference case for both Southeastern Michigan and Northeastern Ohio, with the virus spreading initially at the source of the infection before spreading to surrounding counties. However, the speed at which the infection spreads to surrounding counties is diminished.  

Examining the curves in Figure~\ref{fig:ref_SEIRD} of the different SE(A)IR compartments for the same previous three counties for each region, we find the point in time when peak infections occur for counties that are not the source of the infection to be delated with respect to the reference case. Note that in the reference case, the nature of the network caused a 1 to 2 week delay in peak infection for Genesee and Summit counties with respect to Wayne and Cuyahoga counties, while in this reduced traffic scenario, the delay was more prolonged with 3 to 4 week difference. The overall peak of each of the curve remained similar in profile to what was observed in the reference case. This study shows that while the reduction in commuter traffic may delay the rise of infection, it does not attenuate the severity of the disease. 

% FIGURE 9

% FIGURE 10 
\begin{figure}[h!]
\begin{center}
{\includegraphics[width=\textwidth]{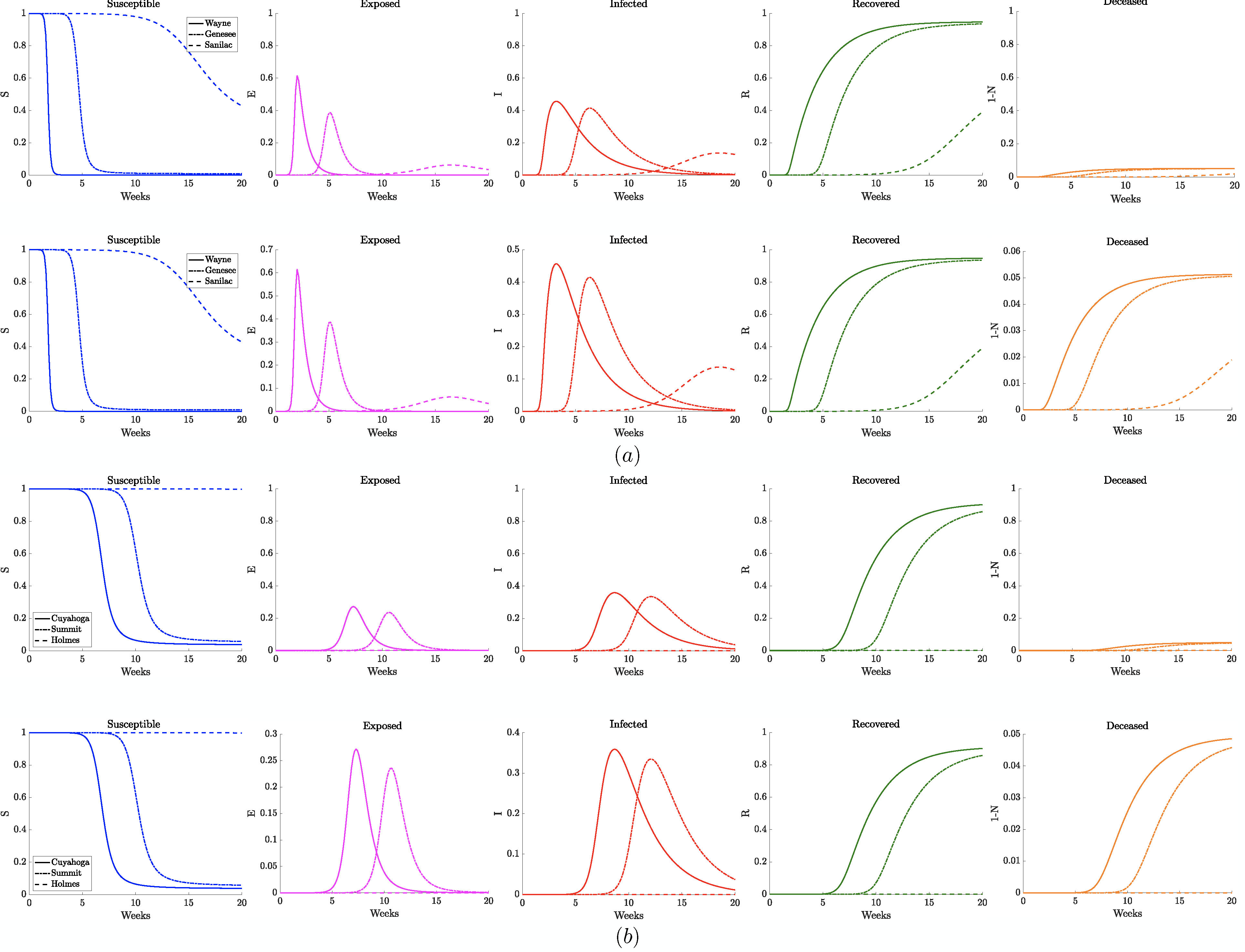}}
\end{center}
\caption{\label{fig:redtraffic_SEIR}  Plots of the relative (left to right) susceptible, exposed, infected, recovered, and deceased populations for the reduced traffic study (Sect.~\ref{sect:redtraffic}) for (a) Wayne, Genesee, and Sanilac counties in southeast Michigan and (b) Cuyahoga, Summit, and Holmes counties in northeast Ohio. The plots on top have the same y-axis scales and the plots on the bottom are with a relative scale for the y-axis.}
\end{figure}

\subsubsection{Varying contact frequency}
\label{sect:varying}

\begin{figure}[h!]
\begin{center}
{\includegraphics[width=\textwidth]{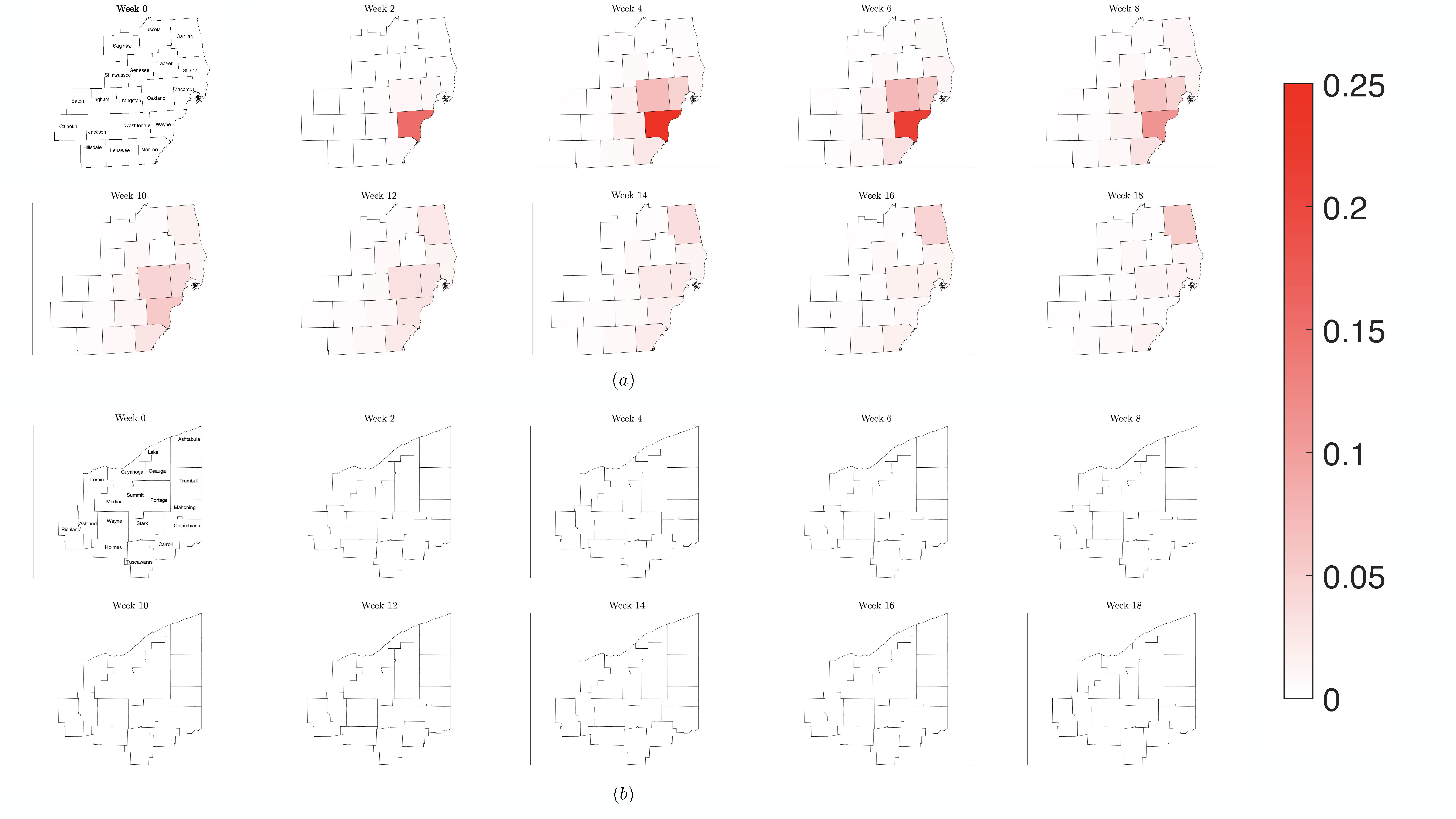}}
\end{center}
\caption{\label{fig:varyr_basemap} Map of the counties in the region of interest of (a) southeast Michigan and (b) northeast Ohio, where the color on the county corresponds to the fraction of the population that is infected for the varying contact frequency study of Sect.~\ref{sect:varying}.}
\end{figure}

\begin{figure}[h!]
\begin{center}
{\includegraphics[width=\textwidth]{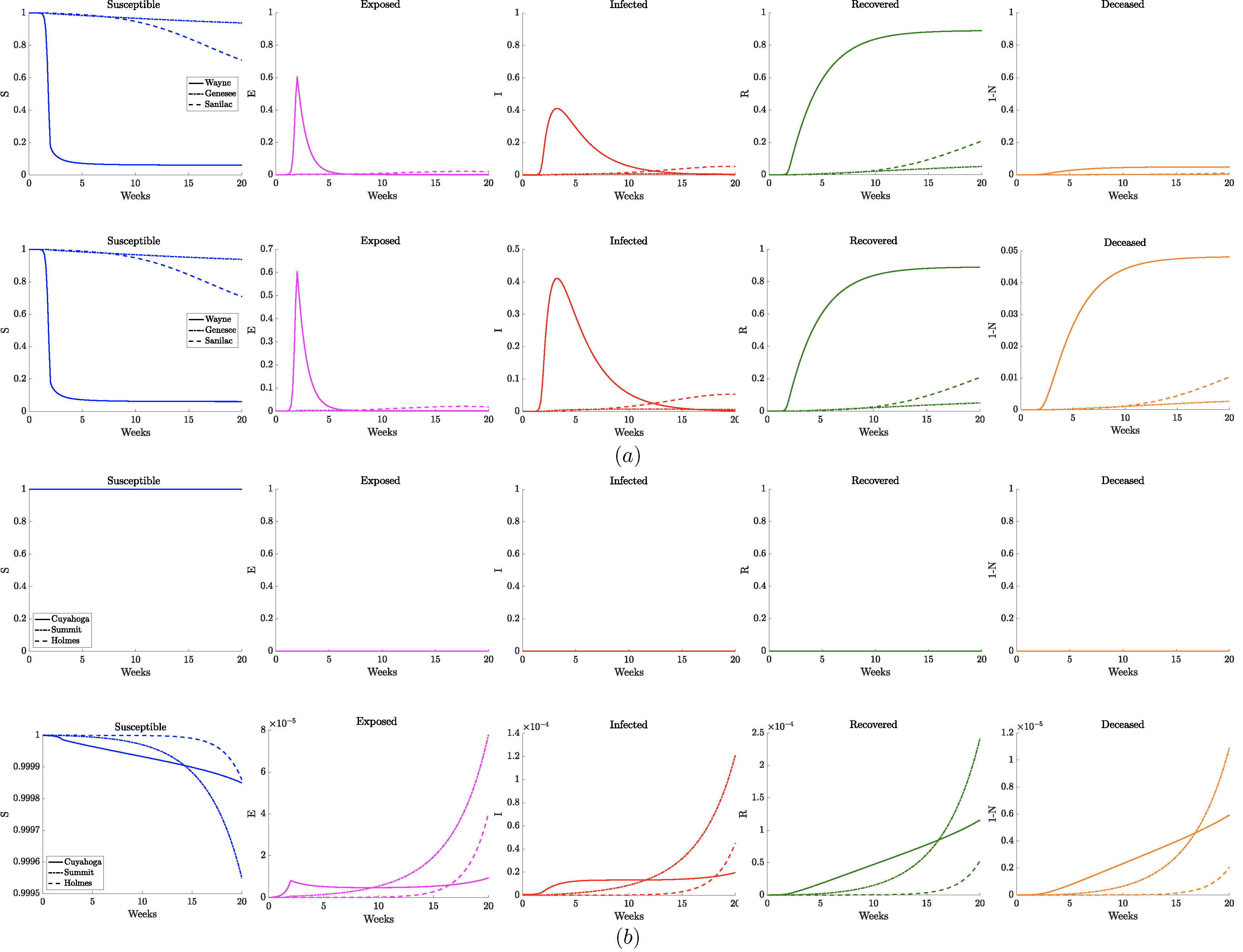}}
\end{center}
\caption{\label{fig:varyr_SEIR} Plots of the relative (left to right) susceptible, exposed, infected, recovered, and deceased populations for the varying contact frequency study (Sect.~\ref{sect:varying}) for (a) Wayne, Genesee, and Sanilac counties in southeast Michigan and (b) Cuyahoga, Summit, and Holmes counties in northeast Ohio. The plots on top have the same y-axis scales and the plots on the bottom are with a relative scale for the y-axis.}
\end{figure}

When estimating the transmission rate in the individual counties using the reported cases, we noted an initial increase before settling to a lower value. This decrease shortly follows the promulgation of stay-at-home orders in both Michigan and Ohio. To account for this, we present a simulation where the frequency of contacts is varied such that it initially has the higher value of the reference case before shifting to a lower contact frequency after 2 weeks. In Figure~\ref{fig:varyr_basemap}, we observe that the virus initially spreads to the three counties that constitute the core of the Detroit metropolitan region before the lower contact frequency regime begins. After the intriduciotn of the lower contact frequency regime, the infection does not spread to other surrounding counties and is contained to the three core counties, as seen in Figure~\ref{fig:varyr_SEIR}. In Ohio, the lower contact frequency regime occurs before the virus is able to gain a foothold in Cuyahoga County. In Figure~\ref{fig:varyr_SEIR}, we notice a small growth in the number of infected, but not enough to result in an appreciable number of infected individuals in the Northeastern Ohio region. These simulations illustrate the pivotal role of contact frequency in determining the trajectory of the disease.

% FIGURE 11

% FIGURE 12

\subsubsection{Testing regimen}
\label{sect:testing}

\begin{figure}[h!]
\begin{center}
{\includegraphics[width=\textwidth]{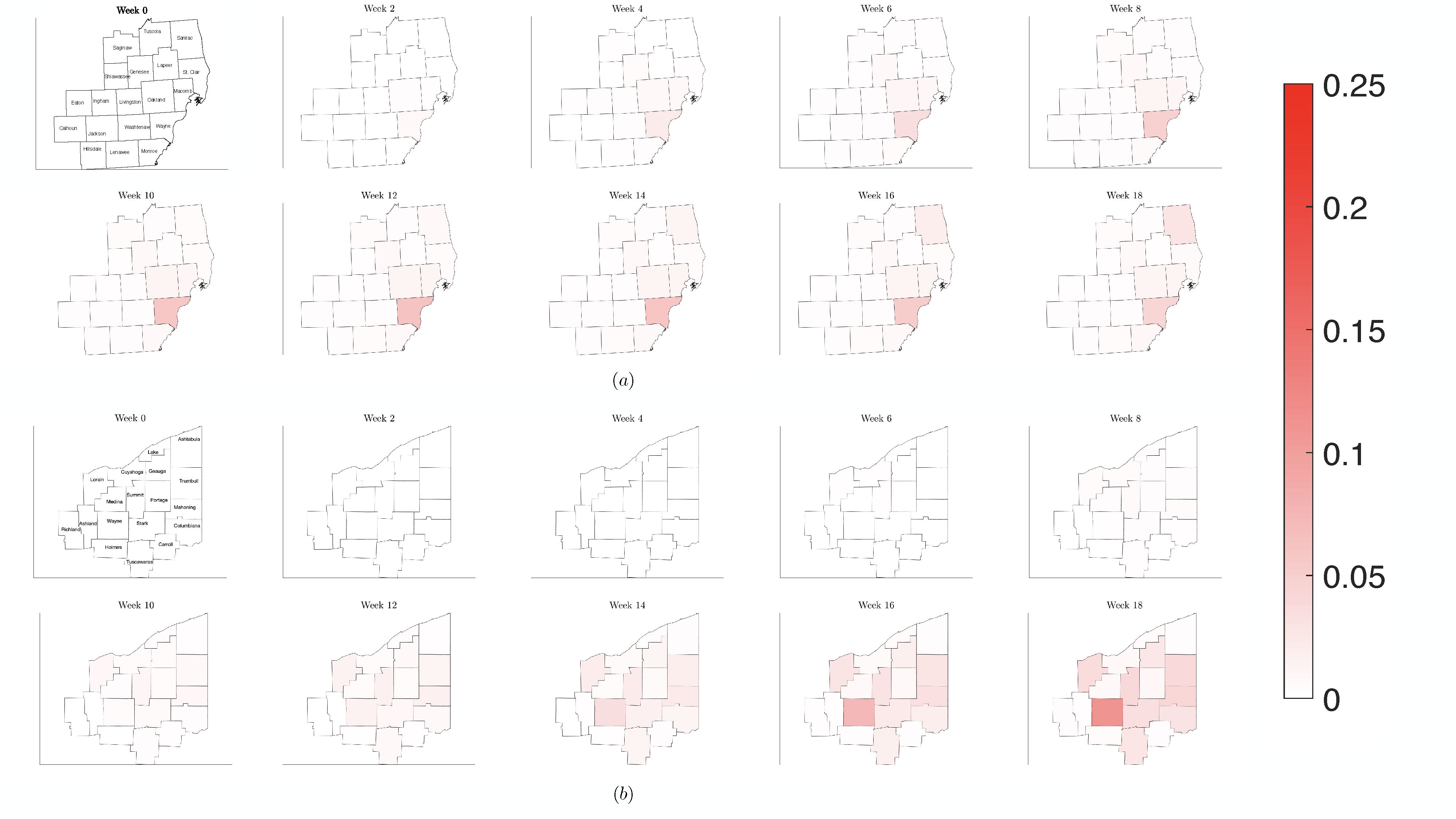}}
\end{center}
\caption{\label{fig:testingr_basemap} Map of the counties in the region of interest of (a) southeast Michigan and (b) northeast Ohio, where the color on the county corresponds to the fraction of the population that is infected for the testing regimen scenario of Sect.~\ref{sect:testing}.}
\end{figure}

\begin{figure}[h!]
\begin{center}
{\includegraphics[width=\textwidth]{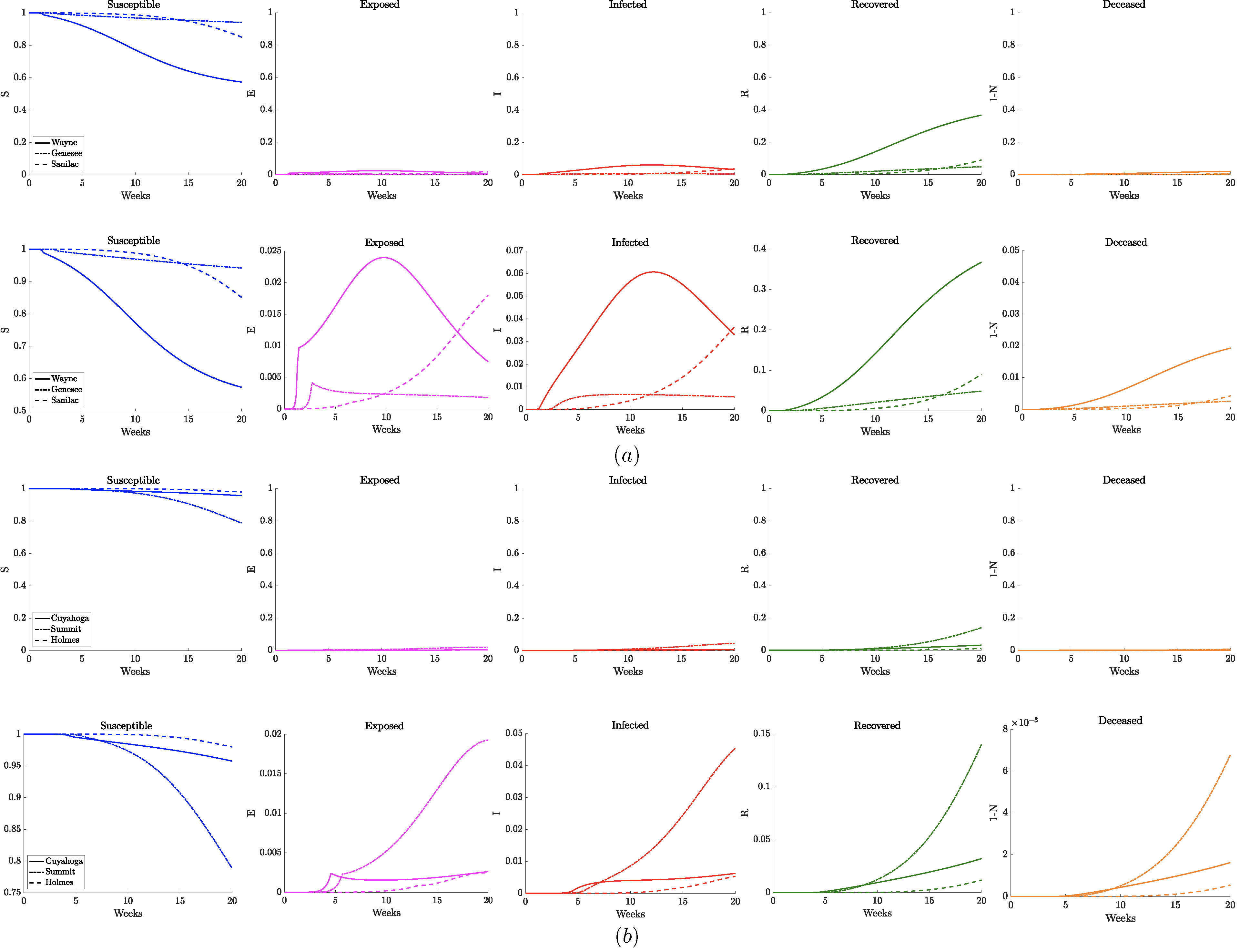}}
\end{center}
\caption{\label{fig:testingr_SEIR} Plots of the relative (left to right) susceptible, exposed, infected, recovered, and deceased populations for the testing regimen scenario (Sect.~\ref{sect:testing}) for (a) Wayne, Genesee, and Sanilac counties in southeast Michigan and (b) Cuyahoga, Summit, and Holmes counties in northeast Ohio. The plots on top have the same y-axis scales and the plots on the bottom are with a relative scale for the y-axis.}
\end{figure}

To illustrate how this network model can be used to evaluate public health policies, we introduce a scenario where a testing regimen is in effect. In this simulation, a higher contact frequency regime is maintained within each county until the infection rate is above 100 per 100,000, at which point the county shifts to a lower contact frequency regime. Applying this regimen to our network model, we observe in Figure~\ref{fig:testingr_basemap} that the number of infected initially increases rapidly for the three counties that make up the Detroit metropolitan region, before the regime shift occurs. This adaptive regimen yields lower overall rates of infection in all three counties and keeps the infection from spreading throughout the region without restricting mobility in any way.  In Figure~\ref{fig:testingr_SEIR}, we find that while the infection does grow slightly, the switch to a lower contact frequency causes a flattening of the curve. In the analogous simulation for Northeast Ohio, the infection increases in Cuyahoga County and spreads to other counties (Figure~\ref{fig:testingr_basemap}), but the overall number of infected people at peak is much lower than for the reference case and reduced traffic study. 

% FIGURE 13 

% FIGURE 14 

\section{Discussion}

The use of mathematical models to study the dynamics of infectious diseases has a long history.  Classical population models \cite{Kermack} have been used extensively to study the spread of epidemics for nearly a century, and the models commonly assume that the populations in the various compartments are homogenous, in the sense that all individuals behave similarly, and well-mixed, i.e., transmission affects all individuals in a compartment at once \cite{anderson1982directly, anderson1985vaccination, keeling2011modeling}. 
These models can be useful to understand the overall dynamics of an epidemics and provide fairly realistic predictions for a homogeneous population, but may not have enough resolution when the population consists of communities with different socio-urban characteristics or demographics.
The need for models that account for the diverse modes of social interaction within each community has been acknowledged for a long time, and the mobility pattern among communities is particularly crucial when trying to forecast the effects of different measures to contain and control the transmission. The concept of metapopulation, intended as a group of spatially separated populations  that have some kind of interaction, was initially introduced for insect pests \cite{levins1969some}  and later used in conjunction with networks to introduce a spatial dimension in modeling transmission of disease.  The network aspect comes from the transfer of individuals among the nodes, and the level of communication between any pair of nodes is determined by the mobility network \cite{perra2011towards}. Information about the mobility network can be gleaned from census data, domestic or international air travel schedules, as relevant for the spatial scale of the model. 

In  \cite{wang2014spatial}, the authors advocate network metapopulation models for describing the spreading of SARS and the outbreaks of A(H1N1) influenza, and A(N7H7), known as avian flu. The common feature of these three epidemics was the speed at which their incidence spread over a wide geographic range: in 2003, SARS-CoV spread from Hong Kong to over 30 countries in 4 continents, before being contained, and in 2009 A(H1N1) spread over 30 countries in 3-4 months. By comparison, SARS-CoV-2 has spread to nearly every country in the world in less than 6 months, following a pattern similar to that of a wild fire. Human recurrent commuting data in metapopulation network models have been used to study changes in contagion processes \cite{balcan2009seasonal, balcan2011phase, balcan2012invasion}. For a comparison of large scale computational approaches to epidemic modeling, in particular agent-based approach versus structured metapopulation models, see  \cite{ajelli2010comparing}.
 Changes in human mobility pattern are often enforced at the outbreak of an epidemic to keep it localized to the original hotspot, however, it has been questioned how effective travel bans really are at containing  a pandemic \cite{hollingsworth2006will, cooper2006delaying, tomba2008simple, bajardi2011human}. To contain COVID-19 pandemic, the measures to control human mobility have varied from the ban of most international flights from affected areas to the near suppression of traffic between communities in regions with high prevalence of infections. In addition, additional changes in mobility have been caused by human reaction to the spread of the epidemics \cite{meloni2011modeling, wang2012safety, wang2013human}.  As reported in \cite{poletti2011effect}, the changing perception of risk did indeed affect the 2009 H1N1 pandemic dynamics.  

The outbreak of SARS-CoV2 in Wuhan, China, and its rapid spread within a few months across Europe and the United States has been closely followed in the hope to find ways to control and contain the pandemic. One important question is where the pandemic will hit next, and how severely the next hotspot will be affected. The geographic pattern followed by the spread of COVID-19 has been rather consistent. As the epidemic moves into a region, the initial hotspots are typically urban centers with large population density and high contact rates, then the infection moves to less crowded communities, where the peak is reached at a later time. The rapidly changing situation and the need of swiftly updating the information based on the inflow of new data underlines the importance of dynamic rather than static models, and the capability of model updating on a daily basis. The proposed Bayesian particle filtering approach combined with a metapopulation network model seeks to address these needs. We refer to \cite{Science_Li} as a  dynamic metapopulation network model with methodology and goals similar to ours, applied to a network of cities in China. The current study concentrated on 18 counties in Northeastern Ohio and 19 counties in Southeastern Michigan, representing a mix of urban, suburban and rural setting. However, as the network is constructed on a basis of publicly available mobility data from the US Census database, the model can be adapted to any county level network.  

As the results show, the time courses of the reproduction number and the transmission rate parameter for the model describing the dynamics of the epidemics in each node, estimated from the daily counts of recorded infections, vary significantly from county to county, however, following a pattern that can be understood in terms of the network connectivity and social distancing measures. In particular, the time courses of the transmission rate in the individual counties clearly demonstrate the effect on this parameter of mitigation measures, mostly in the form of reduced mobility and social distancing. During the observation period included in this study, Ohio started the {\em Stay at home Ohio} program on March 15, 2020, and cancelled the Democratic primary elections, originally scheduled for March 17. A similar {\em Stay at home Michigan} program became effective on March 23, 2023, although the Michigan primary elections had taken place, as scheduled, on March 10. 

\begin{figure}[h!]
\begin{center}
{\includegraphics[width=\textwidth]{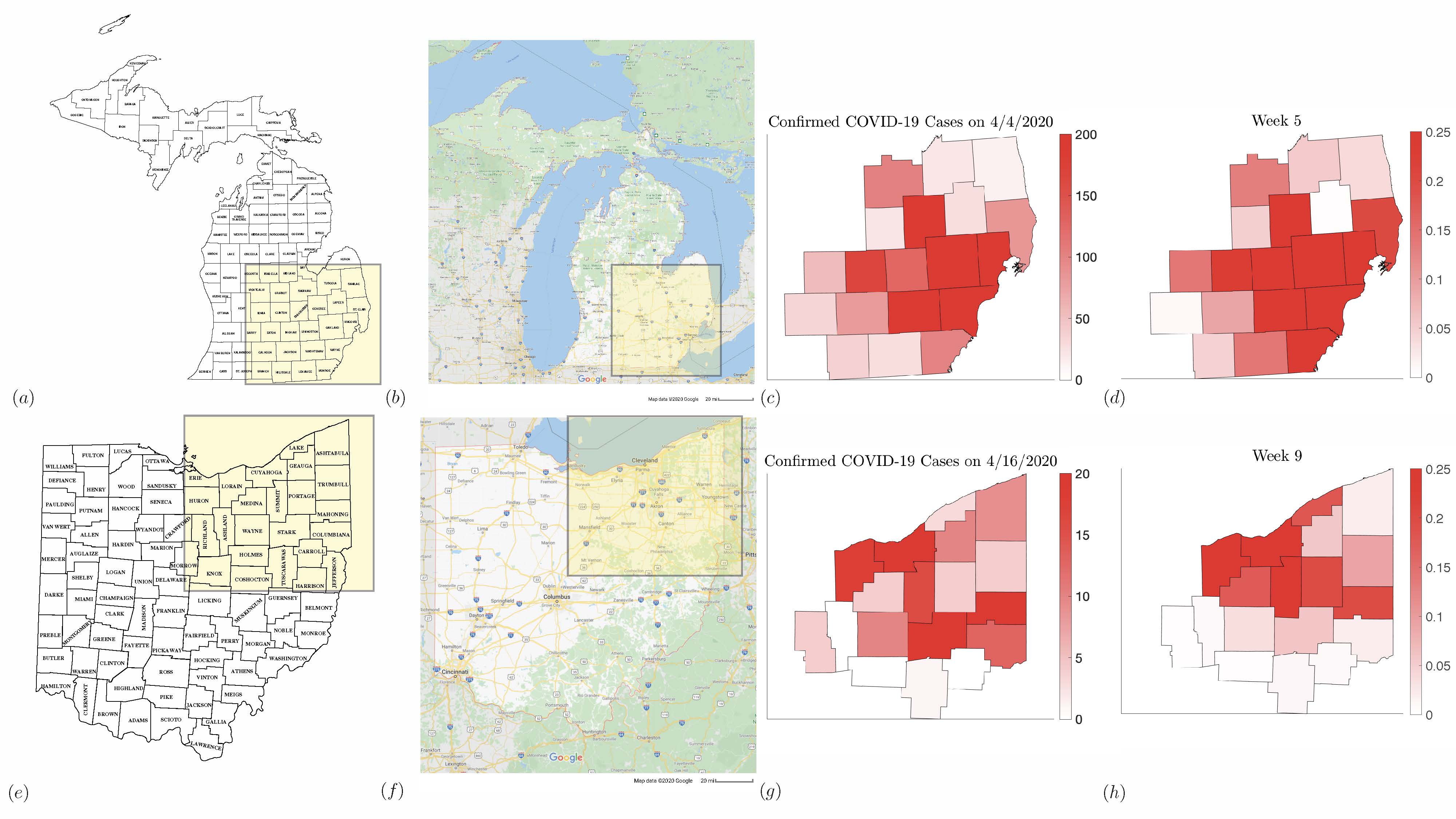}}
\end{center}
\caption{\label{fig:snapshots} Counties in the Michigan (a) and Ohio (e) networks, with the main highways traversing them (panels (b) and (e)). Panels (c) and (f) shows the new number of infected on April 4, 2020 and on April 16, 2020, respectively, with the corresponding model predictions ((d) and (g)) }
\end{figure}

The individual county level models were used to inform the two network models, and it was found that the computed simulations reproduce satisfactorily the observed spreading patterns in both cases, showing the characteristic pattern of the epidemic moving from dense urban centers out, following the highways that directly affect the commuter traffic between the communities.  Fig.~\ref{fig:snapshots} show the block of counties in Michigan (a) and Ohio (e) in the network models and the main highways traversing them (panels (b) and (f)). Panels (c) and (g) show the daily reported number of infections on April 4, 2020 in the Michigan counties, and on April 16, 2020 in the Ohio counties, which are remarkably similar to the model predictions, shown in panel (d) and (h).   The numerical simulations with altered rates of daily contacts clearly demonstrate the effectiveness of social distancing measures in slowing down the epidemics. In particular, two findings are worth highlighting: First, the simulations demonstrate that compared to an non-diversified social distancing measures in all counties, equally effective is a strategy in which social distancing measures are enforced only if the relative frequency of infected individuals exceed a certain threshold. This finding suggests an efficient and economically less burdensome alternative for state-wide mitigation measures, however, it relies heavily on availability of extensive testing. The second finding, in light of the simulations, is the relative inefficiency of travel restrictions. While mobility is undoubtedly the key factor in spreading of epidemics, somewhat counterintuitively, the volume seems to be only a secondary factor, determining how fast the spreading takes place, and not how widely the epidemics spreads. This finding is in line with the discussion in literature \cite{hollingsworth2006will, cooper2006delaying, tomba2008simple, bajardi2011human}.   

% FIGURE 15 

The current version of the simulations does not include demographic information such as age structure of the population that is believed to be an important factor in predicting the severity and outcome of the epidemics for different communities. The demographic data can be introduced in the metapopulation model in a straightforward manner, and preliminary tests of doing this are underway.  Furthermore, a simultaneous and parallel estimation of the parameters of connected communities will be part of the future work. As demonstrated in earlier articles by the authors \cite{arnold2014vectorized}, the particle filtering technique is particularly well suited for parallel computing.

\section*{Conflict of Interest Statement}

The authors declare that the research was conducted in the absence of any commercial or financial relationships that could be construed as a potential conflict of interest.

\section*{Author Contributions}

All authors participated in the design of the models, methodology, and experiments, and assessment of their relevance.
DC, APH and ES contributed the computer coding and testing. All authors participated in the writing of the article.

\section*{Funding}
The work of DC was partly supported by NSF grants DMS-1522334 and DMS-1951446, and the work of ES was partly supported by NSF grant DMS-1714617.

%\section*{Acknowledgments}
%This is a short text to acknowledge the contributions of specific colleagues, institutions, or agencies that aided the efforts of the authors.

%\section*{Supplemental Data}
% \href{http://home.frontiersin.org/about/author-guidelines#SupplementaryMaterial}{Supplementary Material} %should be uploaded separately on submission, if there are Supplementary Figures, please include the caption in the same file as the figure. LaTeX Supplementary Material templates can be found in the Frontiers LaTeX folder.

\section*{Data Availability Statement}
The datasets utilized for this study can be found in the USAfacts webpage \begin{verbatim}{https://usafacts.org/} \end{verbatim}.
% Please see the availability of data guidelines for more information, at https://www.frontiersin.org/about/author-guidelines#AvailabilityofData

%\bibliographystyle{frontiersinSCNS_ENG_HUMS} % for Science, Engineering and Humanities and Social Sciences articles, for Humanities and Social Sciences articles please include page numbers in the in-text citations
%\bibliographystyle{frontiersinHLTH&FPHY} % for Health, Physics and Mathematics articles
\bibliographystyle{unsrt}
\bibliography{References}

\end{document}